\documentclass[aps,prd,twocolumn,epsf,nofootinbib,superscriptaddress,showpacs]{revtex4}

\usepackage{graphicx,here}
\usepackage{amsmath,amssymb,latexsym}
\usepackage{bm}
\usepackage{url}

\def\lsim{\mathrel{\raise.3ex\hbox{$<$\kern-.75em\lower1ex\hbox{$\sim$}}}}
\def\gsim{\mathrel{\raise.3ex\hbox{$>$\kern-.75em\lower1ex\hbox{$\sim$}}}}

\def\lbldef#1#2{\expandafter\gdef\csname #1\endcsname {#2}}

\def\href#1#2{#2}

\newcommand{\bwide}{\begin{widetext}}
\newcommand{\ewide}{\end{widetext}}
\newcommand{\beq}[1]{\begin{equation} \label{(#1)}}
\newcommand{\eeq}{\end{equation}}
\newcommand{\ba}[1]{\begin{eqnarray} \label{(#1)}}
\newcommand{\ea}{\end{eqnarray}}

\begin{document}


\title{Searching For Dark Matter Subhalos In the Fermi-LAT Second Source Catalog}

\author{Alexander V.~Belikov}
\affiliation{Institut d'Astrophysique de Paris, UMR 7095, CNRS, UPMC Univ. Paris
06, 98 bis boulevard Arago, 75014 Paris, France}
\author{Matthew R.~Buckley}
\affiliation{Center for Particle Astrophysics, Fermi National Accelerator Laboratory, Batavia, IL 60510, USA}
\author{Dan Hooper}
\affiliation{Center for Particle Astrophysics, Fermi National Accelerator Laboratory, Batavia, IL 60510, USA}
\affiliation{Department of Astronomy and Astrophysics, The University of Chicago, Chicago, IL  60637, USA} 

\begin{abstract}

The dark matter halo of the Milky Way is expected to contain an abundance of smaller subhalos. These subhalos can be dense and produce potentially observable fluxes of gamma rays. In this paper, we search for dark matter subhalo candidates among the sources in the Fermi-LAT Second Source Catalog which are not currently identified or associated with counterparts at other wavelengths. Of the nine high-significance, high-latitude ($|b|>60^{\circ}$), non-variable, unidentified sources contained in this catalog, only one or two are compatible with the spectrum of a dark matter particle heavier than approximately $50$--$100$ GeV. The majority of these nine sources, however, feature a spectrum that is compatible with that predicted from a lighter ($\sim$5-40 GeV) dark matter particle. This population is consistent with the number of observable subhalos predicted for a dark matter candidate in this mass range and with an annihilation cross section of a simple thermal relic ($\sigma v\sim3\times 10^{-26}$ cm$^3$/s). Observations in the direction of these sources at other wavelengths will be necessary to either reveal their astrophysical nature (as blazars or other active galactic nuclei, for example), or to further support the possibility that they are dark matter subhalos by failing to detect any non-gamma ray counterpart.

\end{abstract}

\pacs{95.35.+d;07.85.-m;98.70.Rz; FERMILAB-PUB-11-606-A}

\maketitle

\section{Introduction}

Numerical simulations have revealed that dark matter structures form hierarchically, with small halos merging to form increasingly massive halos. As a consequence of this process, dark matter halos (such as the one hosting the Milky Way) are expected to contain many subhalos, ranging from relatively large dwarf spheroidal galaxies ($M \gsim 10^7 M_{\odot}$) to objects with masses as small as $10^{-3}-10^{-8} M_{\odot}$.

Despite their ubiquity, dark matter subhalos are very difficult to detect or otherwise observe. Subhalos that are less massive than dwarf galaxies are not expected to contain significant quantities or stars or gas, making them invisible to conventional telescopes. Dark matter annihilations taking place in such subhalos, however, could potentially produce observable fluxes of gamma rays. If the dark matter particles are very heavy ($m_{\rm DM} \gsim\,$TeV), such gamma rays could be observed by ground based gamma ray telescopes~\cite{acts}. For less massive dark matter particles ($m_{\rm DM}\sim$ GeV-TeV), the space-based Fermi Gamma Ray Space Telescope (FGST) provides the greatest prospects for detecting the products of dark matter annihilations in subhalos~\cite{Buckley:2010vg,Kuhlen:2008aw,Zechlin:2011wa,Mirabal:2010ny}.

To the FGST, relatively large ($\sim$$10^2$--$10^6 M_{\odot}$) and nearby ($\sim$$0.01$--$1$ kpc) subhalos could appear as bright and approximately point-like gamma ray sources, without detectable counterparts in other wavelengths. For reasonable estimates of the number of subhalos in the Milky Way, the dark matter distribution within those subhalos, and characteristics of the dark matter particles themselves, one is led to expect on the order of 0.1-10 subhalos to be observable by the FGST. In particular, a typical dark matter candidate with a mass of 10-100 GeV and an annihilation cross section of $\sigma v \sim 3 \times 10^{-26}$ cm$^3$/s is expected to provide on the order of a few subhalos that are observable at or above the $5\sigma$ level. 

The recently published Second Fermi-LAT Source Catalog (2FGL)~\cite{2catalog} contains many unidentified sources which could plausibly be dark matter subhalos. In particular, of among the 1873 gamma ray sources contained in this catalog, 576 have not been associated with counterparts at other wavelengths or been otherwise identified. From among this sample, 397 have been detected at greater than 5$\sigma$ significance, and do not exhibit discernible variability. In this paper, we study this sample of unidentified gamma ray sources and consider whether any significant fraction of them might be the result of dark matter annihilations taking place in nearby subhalos.

The remainder of the article is structured as follows. In Sec.~\ref{theory}, we describe our calculation of the number of dark matter subhalos observable by the FGST. In Sec.~\ref{catalogsec}, we study the contents of the Second Fermi-LAT Source Catalog (2FGL) and attempt to identify dark matter subhalo candidates. While no conclusive identifications can be made at this time, we find that many or most of the unidentified high-latitude ($|b|>60^{\circ}$) sources could be dark matter subhalos if the dark matter is relatively light ($m_{\rm DM} \sim 5$--$40$ GeV) and possesses an annihilation cross section on the order of that predicted for a simple thermal relic ($\sim 3\times 10^{-26}$ cm$^3$/s). In Sec.~\ref{gc}, we discuss this result within the context of evidence for $\sim$10 GeV dark matter particles from the Galactic Center, non-thermal radio filaments, and the Milky Way's microwave haze. In Sec.~\ref{conclusions} we summarize our results and conclusions.

\section{Gamma Rays From Nearby Dark Matter Subhalos}
\label{theory}

Numerical simulations of the formation and evolution of structure have revealed that dark matter halos and subhalos form with a distribution of masses given by $dN_n/dM_h \propto M_h^{-2}$, extending down to a minimum mass related to the microscopic properties of the dark matter particle~\cite{smallest}. For a Milky Way sized halo, on the order of 5-10\% (50\%) of the total mass is expected to be found in $10^7$-$10^{10} \, M_{\odot}$ ($10^{-6}$-$10^{10} \, M_{\odot}$) subhalos~\cite{norm}.  For a minimum subhalo mass of one Earth mass, this normalization corresponds to a total of more than $10^{16}$ subhalos within the halo of the Milky Way, for a number density of approximately $\sim$$10^2$ subhalos per cubic parsec in the local neighborhood of the Galaxy. 

The luminosity of gamma rays from dark matter annihilations in a given subhalo depends strongly on how the dark matter is distributed. Numerical simulations find that dark matter halos (and subhalos) possess density profiles which approximately take the form~\cite{1pt2}:
\begin{equation}
\rho(r) \propto \frac{1}{(r/R_s)^{\gamma}\, [1+(r/R_s)]^2},
\label{1pt2}
\end{equation}
where $R_s$ is the scale radius and $\gamma$ is a parameter which dictates the inner slope of the profile. In the case of the traditional Navarro, Frenk and White (NFW) profile, $\gamma=1$, while more recent results from the Via Lactea collaboration favor somewhat larger values, $\gamma \approx 1.2$~\cite{Diemand:2009bm}. The Aquarius Project, in contrast, has reported profiles of varying slope which are somewhat less steep in the innermost regions~\cite{shallow}. Subhalos with such a profile produce similar gamma ray fluxes to those with an inner slope of $\gamma \sim 1.1$ (see Table~1 of Ref.~\cite{Buckley:2010vg}, for example).


The concentration of a halo is defined as the ratio of its virial radius to its scale radius. Halos with larger concentrations have more of their mass located in their inner volumes, leading to higher annihilation rates. To estimate the concentration of a halo of a given mass, we use the analytic model of Bullock {\it et al.}~\cite{bullock}. This model estimates a concentration of 21 (27) for a subhalo of mass $10^{7} M_{\odot}$ ($10^{5} M_{\odot}$). For very massive halos, the results of numerical simulations are in good agreement with this model. Considerable halo-to-halo variation in the concentration and shape of subhalo profiles has been observed in numerical simulations. The model of Bullock {\it et al.}, for example, only provides a measure of the average concentration of a subhalo of a given mass. The probability of a halo having a given concentration can be modelled by a log-normal distribution, with a dispersion of $\sigma_c\approx 0.24$~\cite{bullock}. We include this variation in our calculations, which in most cases roughly doubles the number of observable subhalos (see also Ref.~\cite{Pieri:2007ir}).

Subhalos in the local volume of the Milky Way are likely to have had a large fraction of their mass stripped through tidal interactions with other halos and stars. This primarily impacts a subhalo's outer density profile, leaving a dense and tightly bound inner cusp intact~\cite{disruption}. As the innermost regions of halos dominate the overall annihilation rate, the precise fraction of mass that is lost only modestly impacts the resulting gamma ray luminosity.


The rate of gamma rays produced from dark matter annihilations taking place in a nearby subhalo is given by
\begin{equation}
L_{\gamma}=\frac{\sigma v}{2 m^2_{\rm DM}} \int \frac{dN_{\gamma}}{dE_{\gamma}} dE_{\gamma} \int \rho^2 dV,
\end{equation}
where $\sigma v$ and $m_{\rm DM}$ are the annihilation cross section and mass of the dark matter particle, respectively, and the second integral is performed over the volume of the subhalo. $dN_{\gamma}/dE_{\gamma}$ is the spectrum of gamma rays produced per dark matter annihilation, which depends on the dominant annihilation channel(s) and on the mass of the dark matter particle (we use PYTHIA 6~\cite{pythia} to calculate the gamma ray spectrum). 


In order for the gamma ray annihilation products from a subhalo to constitute a source that could potentially appear within the FGST second source catalog, the subhalo must be both sufficiently bright and sufficiently compact to mimic an approximate point source. To estimate the number of events from a subhalo observed by FGST, we multiply the gamma ray flux by an effective area of 6800 cm$^2$, and a coverage of 20\% of the sky at any given time. Although the detectability of a given gamma ray source depends somewhat on its spectral shape and its location in the sky, we can roughly estimate how bright a given subhalo must be to be detectable at high significance by FGST. In particular, the diffuse gamma ray flux measured by FGST generates approximately 20 events per year per square degree above 1 GeV over Galactic latitudes of $|b| > 60^{\circ}$, and about 60-70 events per year per square degree above 1 GeV over Galactic latitudes of $10^{\circ}<|b|<20^{\circ}$. In these two regions of sky, we estimate that $5\sqrt{20} \approx 20$ or $5\sqrt{60} \approx 40$ signal events per year above 1 GeV would be required in order for a subhalo to be potentially discovered with 5$\sigma$ significance. Based on this estimate, we conservatively classify any subhalo that produces more than 50 events above 1 GeV per year at FGST to be potentially detectable. We further support this estimate by noting that the least significant sources contained in the second Fermi source catalog (2FGL) are of approximately this brightness~\cite{2catalog}.

\begin{figure*}[t]
\begin{center}
{\includegraphics[angle=0,width=0.9\linewidth]{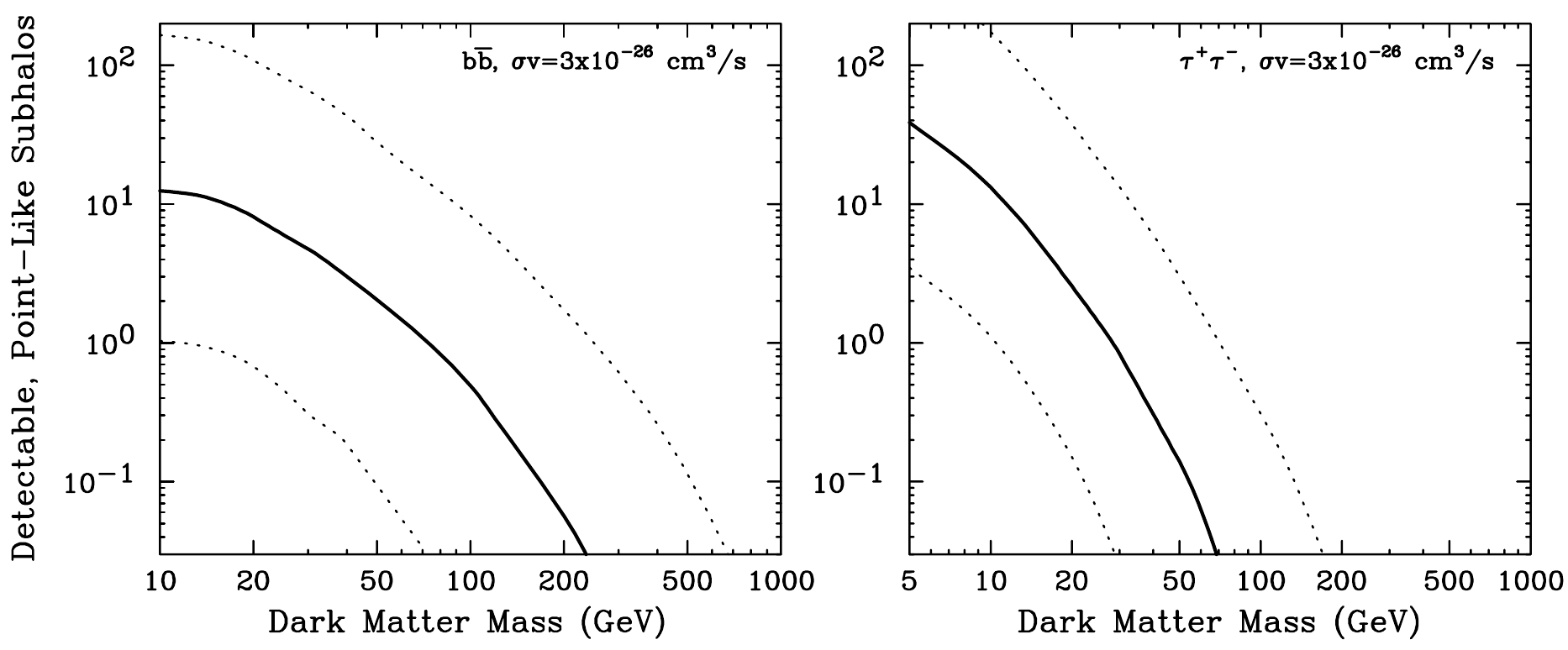}}\\
{\includegraphics[angle=0,width=0.9\linewidth]{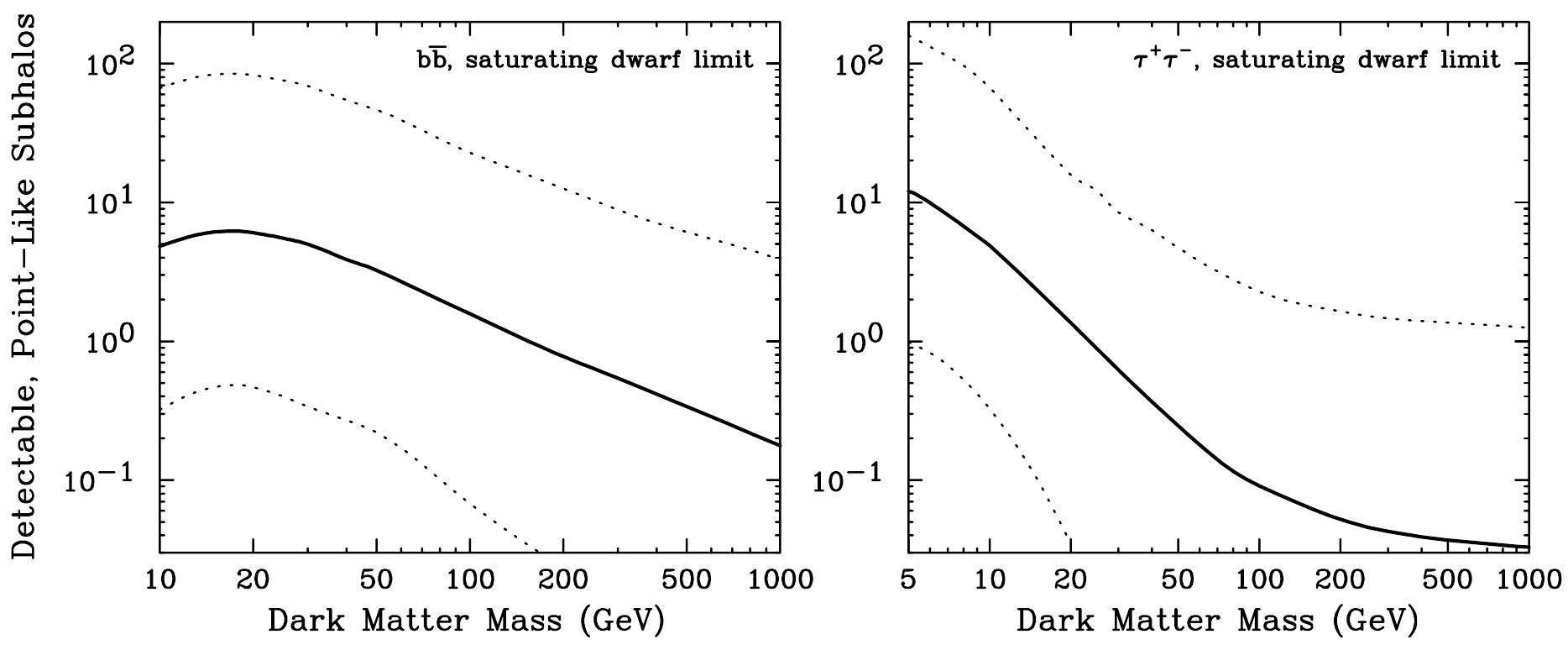}}
\vspace{-0.3cm}
\caption{The number of point-like subhalos potentially detectable by the Fermi Gamma Ray Space Telescope as a function of the dark matter's mass, and assuming an annihilation cross section of $\sigma v = 3\times 10^{-26}$ cm$^3$/s (top), or one which saturates the current constraints based on observations of dwarf galaxies (bottom)~\cite{dwarfs}. To qualify as detectable and point-like, we require a subhalo to produce more than 50 events above 1 GeV per year at FGST, and require that more than 95\% of those photons be concentrated within a radius of $2^{\circ}$. We show results for two dark matter annihilation channels ($b\bar{b}$ and $\tau^+ \tau^-$). Results using our central (optimistic/pessimistic) parameter estimates are shown as solid (dotted) lines. See text for more details}
\label{number}
\end{center}
\end{figure*}

In order to be contained in the 2FGL, a given source must be point-like or possess only a minimal angular extent.\footnote{Although 12 sources in the 2FGL have been identified as extended, none of the unidentified sources are presently inconsistent with being point-like~\cite{2catalog}.} In other words, the angular extent of a subhalo must not be much larger than the telescope's point spread function. In this work, to be considered sufficiently point-like, we require that 95\% of the photons from a subhalo come from within a 2$^{\circ}$ radius (approximately the 95\% containment angle for 1 GeV photons at the FGST).

In calculating estimates for the number of dark matter subhalos to appear in the 2FGL, we adopt a range of physical assumptions. For our central estimates, we assume that the outer 99\% of the original halo's mass is lost to tidal stripping, adopt an inner slope of $\gamma=1.1$ for the density profile, and integrate down to substructure masses of 10$^{-6}\, M_{\odot}$ (corresponding to a boost factor\footnote{By ``boost factor'', we mean the overall enhancement to the annihilation rate resulting from sub-subhalos existing within the larger subhalo.} of 1.75). We also consider a more optimistic (95\% mass loss, $\gamma=1.2$, boost factor=2.8) and pessimistic (99.5\% mass loss, $\gamma=1.0$, no boost factor) set of assumptions. In each case, we adopt concentrations given by the Bullock {\it et al.} model~\cite{bullock}. We consider this range of possibilities to realistically reflect the uncertainites involved in this calculation.

In the upper frames of Fig.~\ref{number}, we show our estimates for the number of dark matter subhalos appearing as unidentified sources in the 2FGL.  Here, we have adopted an annihilation cross section of $\sigma v =  3 \times 10^{-26}$ cm$^3$/s, as predicted for a simple thermal relic. We show results for a range of masses, and for annihilations to $b\bar{b}$ (left) and $\tau^+ \tau^-$ (right).\footnote{Annihilations to lighter quarks or gauge bosons produce gamma ray spectra similar to the $b\bar{b}$ case. We do not consider annihilations to electrons or muons here, as they lead to relatively few gamma rays (through final state radiation).} In each frame, the solid line denotes the result found using our central estimates, while the dotted lines use our more optimistic and pessimistic assumptions. Our results are in good agreement with those presented elsewhere, such as in Ref.~\cite{Pieri:2007ir} for example.

In some dark matter models (with non-thermal histories, or Sommerfeld enhancements, for example), the annihilation cross section can be larger than $\sigma v =  3 \times 10^{-26}$ cm$^3$/s, potentially leading to a larger number of detectable subhalos. In the lower frames of Fig.~\ref{number}, we estimate the number of observable subhalos using an annihilation cross section equal to the upper limits placed by the Fermi collaboration, based on the observation of dwarf galaxies~\cite{dwarfs} (constraints that are at least as stringent have also been derived from observations of the Galactic Center~\cite{Hooper:2011ti}). These constraints in turn, can be translated into a maximum number of subhalos that could be potentially detectable by Fermi. A 100 GeV dark matter particle annihilating to $b\bar{b}$, for example, is expected to lead to no more than 1.6 observable subhalos across the sky using our central parameter estimates, and no more than 23 observable subhalos using our optimistic parameters. In each case, most of the potentially detectable subhalos and nearby (within roughly one kiloparsec) and have fairly large masses, $M \gsim 10^4 M_{\odot}$ (after accounting for mass loss due to tidal stripping).

\begin{figure*}[t]
\begin{center}
{\includegraphics[angle=0,width=0.46\linewidth]{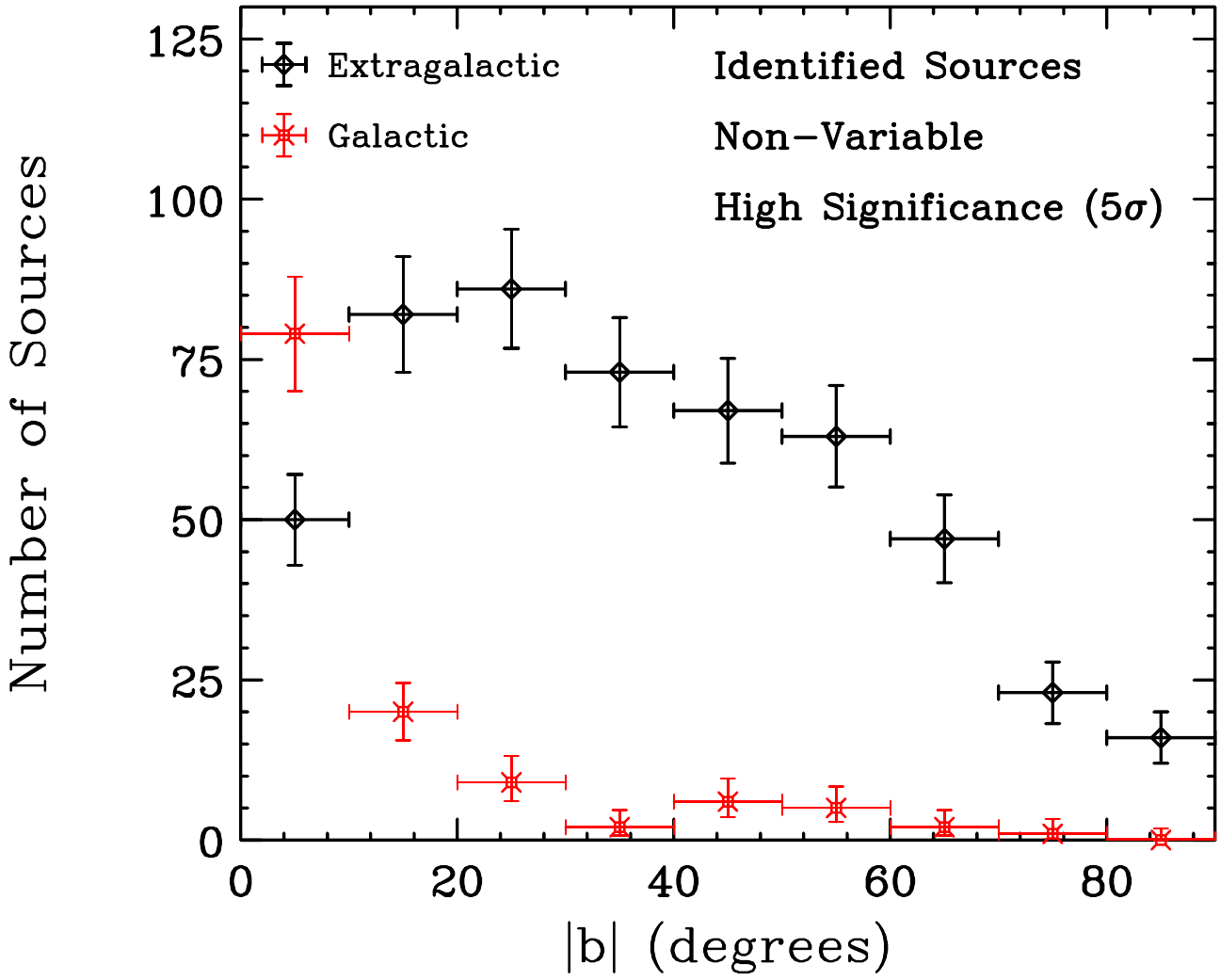}}
\hspace{0.02\linewidth}
{\includegraphics[angle=0,width=0.46\linewidth]{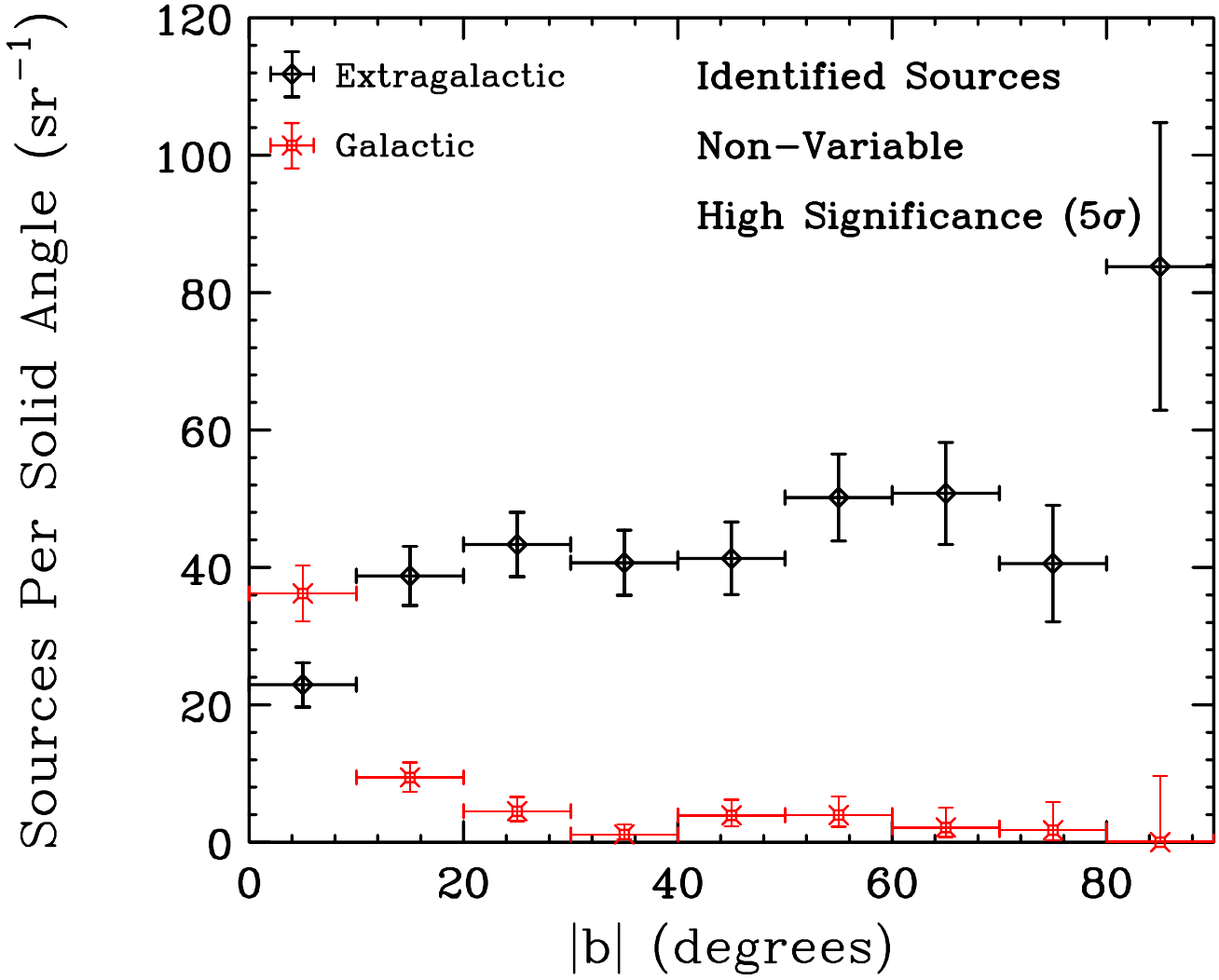}}\\
{\includegraphics[angle=0,width=0.46\linewidth]{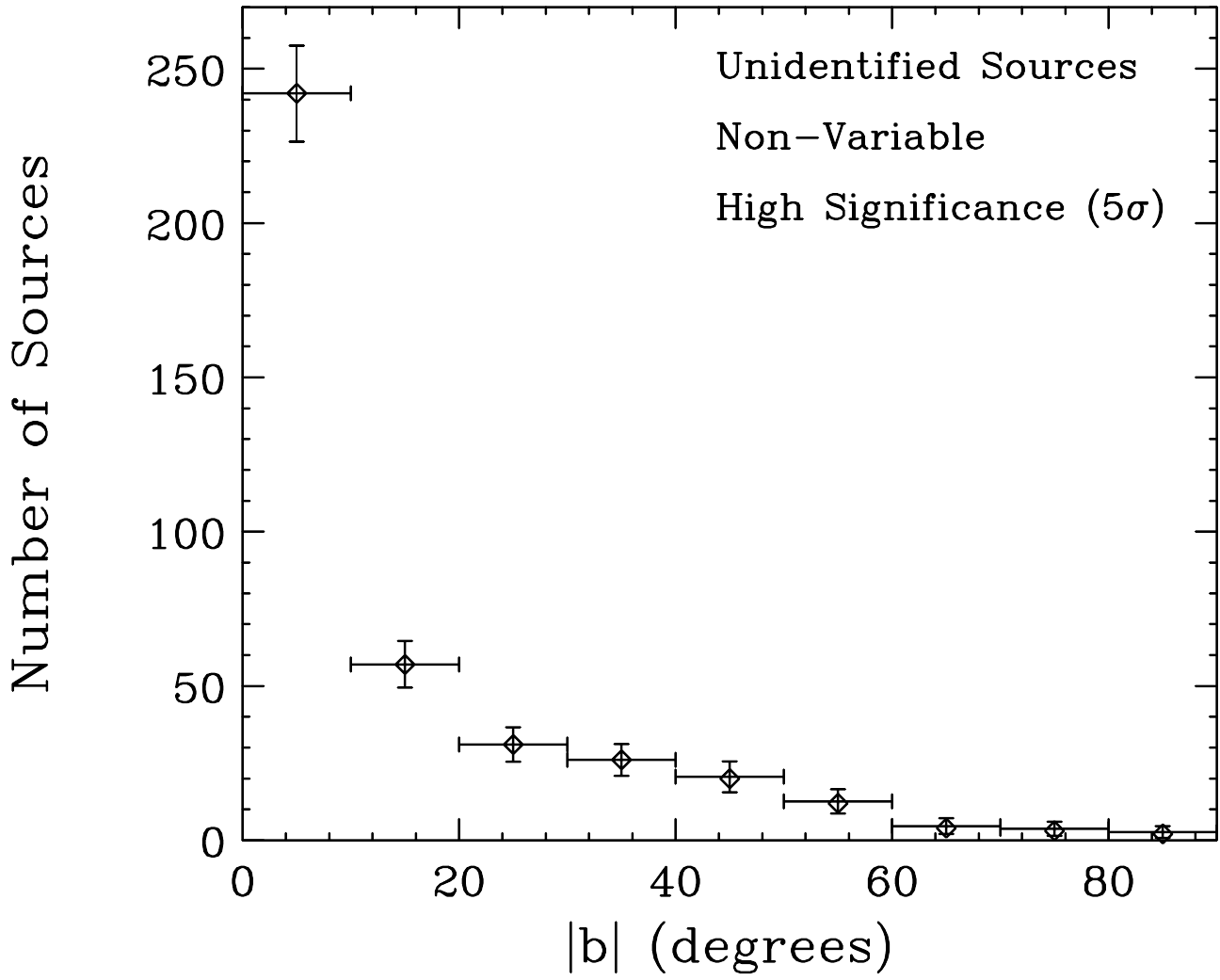}}
\hspace{0.02\linewidth}
{\includegraphics[angle=0,width=0.46\linewidth]{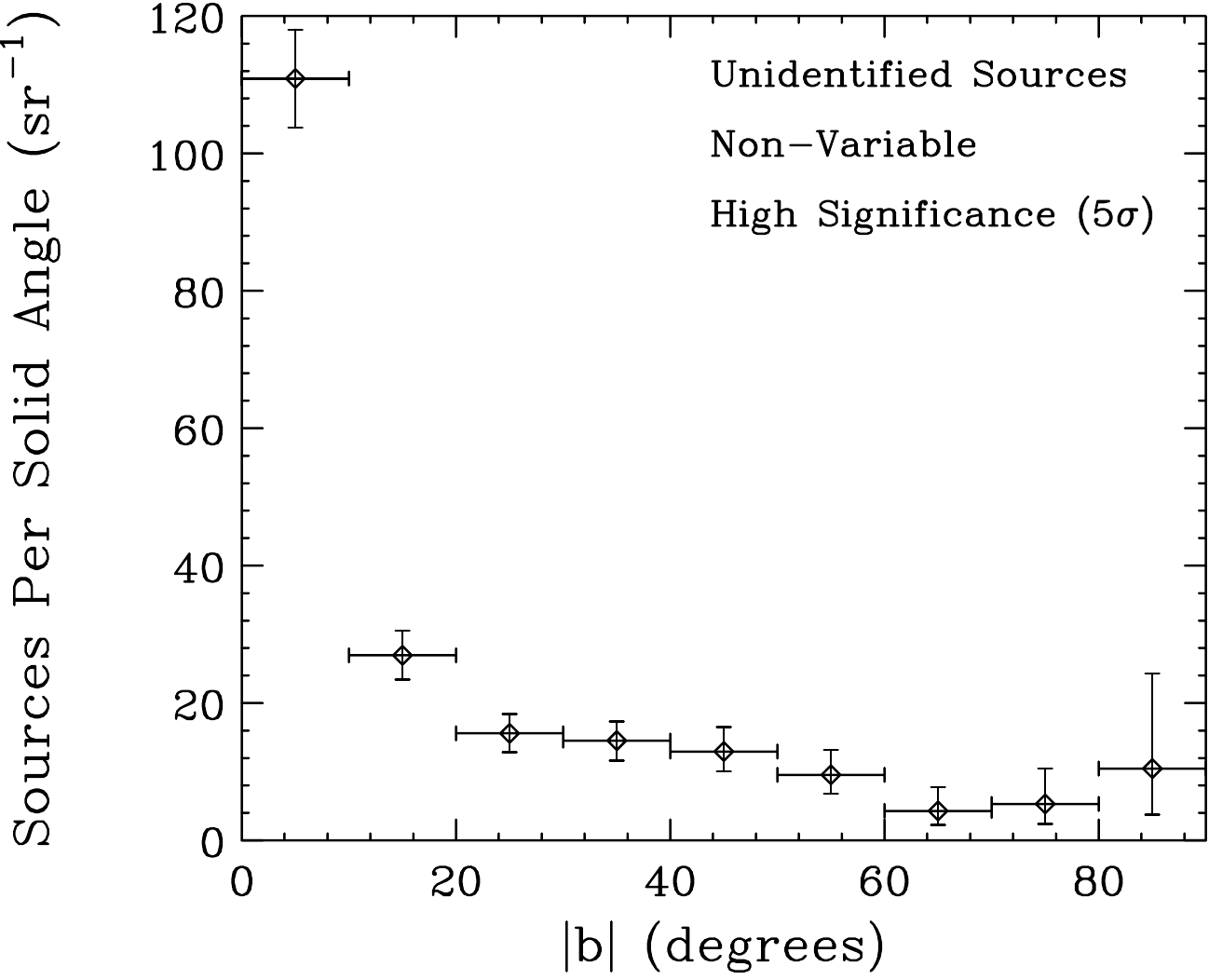}}
\caption{The distribution in Galactic latitude of identified (top) and unidentified (bottom) sources in the second Fermi-LAT source catalog. Only non-variable sources which have been detected at 5$\sigma$ or greater significance have been included. See text for more details.}
\label{catalog}
\end{center}
\end{figure*}

Based on these estimates, we conclude that if the dark matter particles are relatively light, up to $\sim$10 (or up to $\sim$100 using optimistic estimates) subhalos could appear as unidentified gamma ray sources in the 2FGL. In the following section, we discuss the characteristics of this catalog and the prospects of identifying dark matter subhalos from among its constituents.


\section{Subhalos in the Fermi-LAT Second Source Catalog}
\label{catalogsec}

\begin{figure*}[t]
\begin{center}
{\includegraphics[angle=0,width=0.92\linewidth]{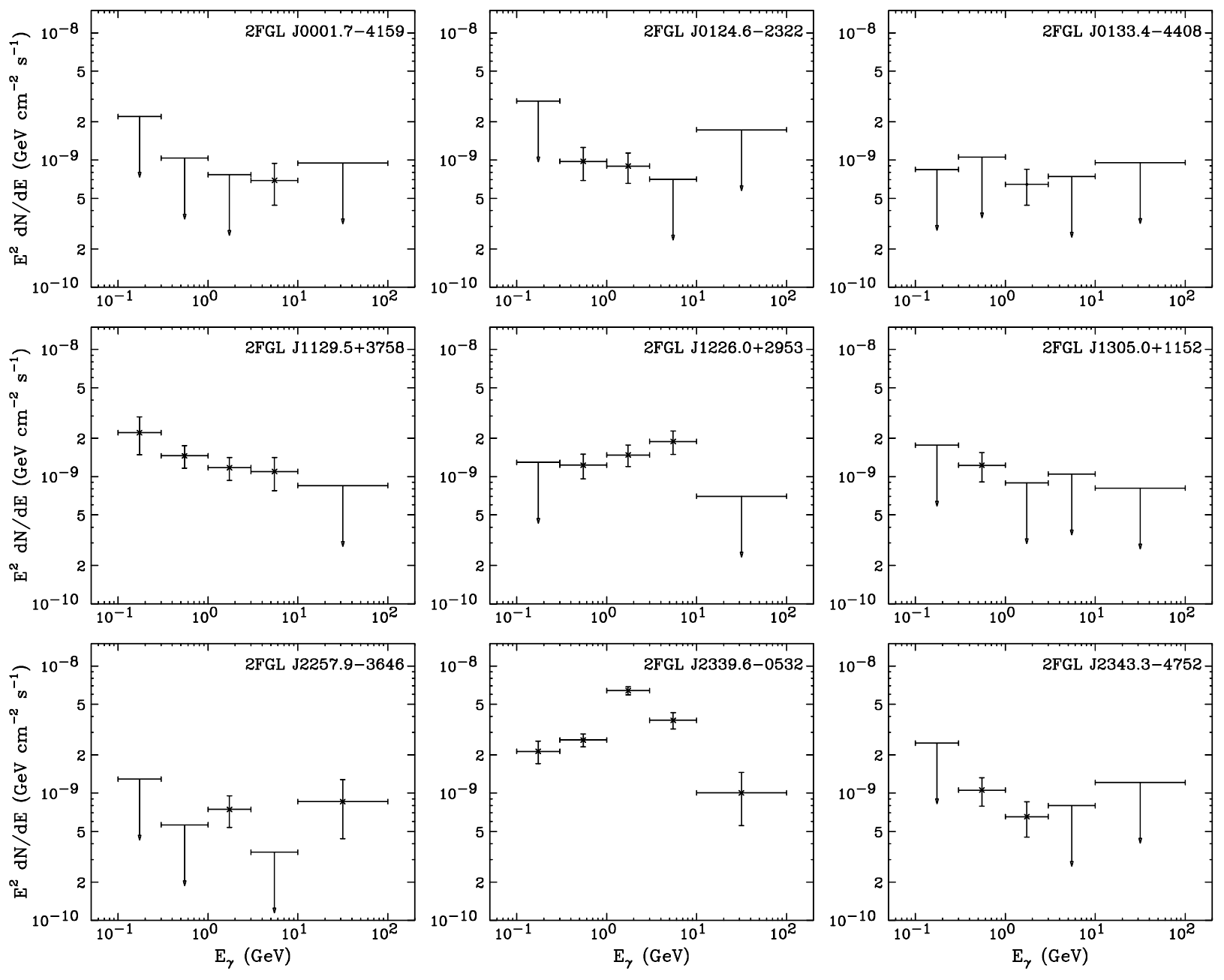}}
\caption{Spectra of the nine high-latitude ($|b|>60^{\circ}$), high-significance ($>5\sigma$), non-variable, unidentified sources contained in the Fermi Second Source Catalog (2FGL).}
\label{nine}
\end{center}
\end{figure*}

\begin{figure*}[t]
\begin{center}
{\includegraphics[angle=0,width=0.46\linewidth]{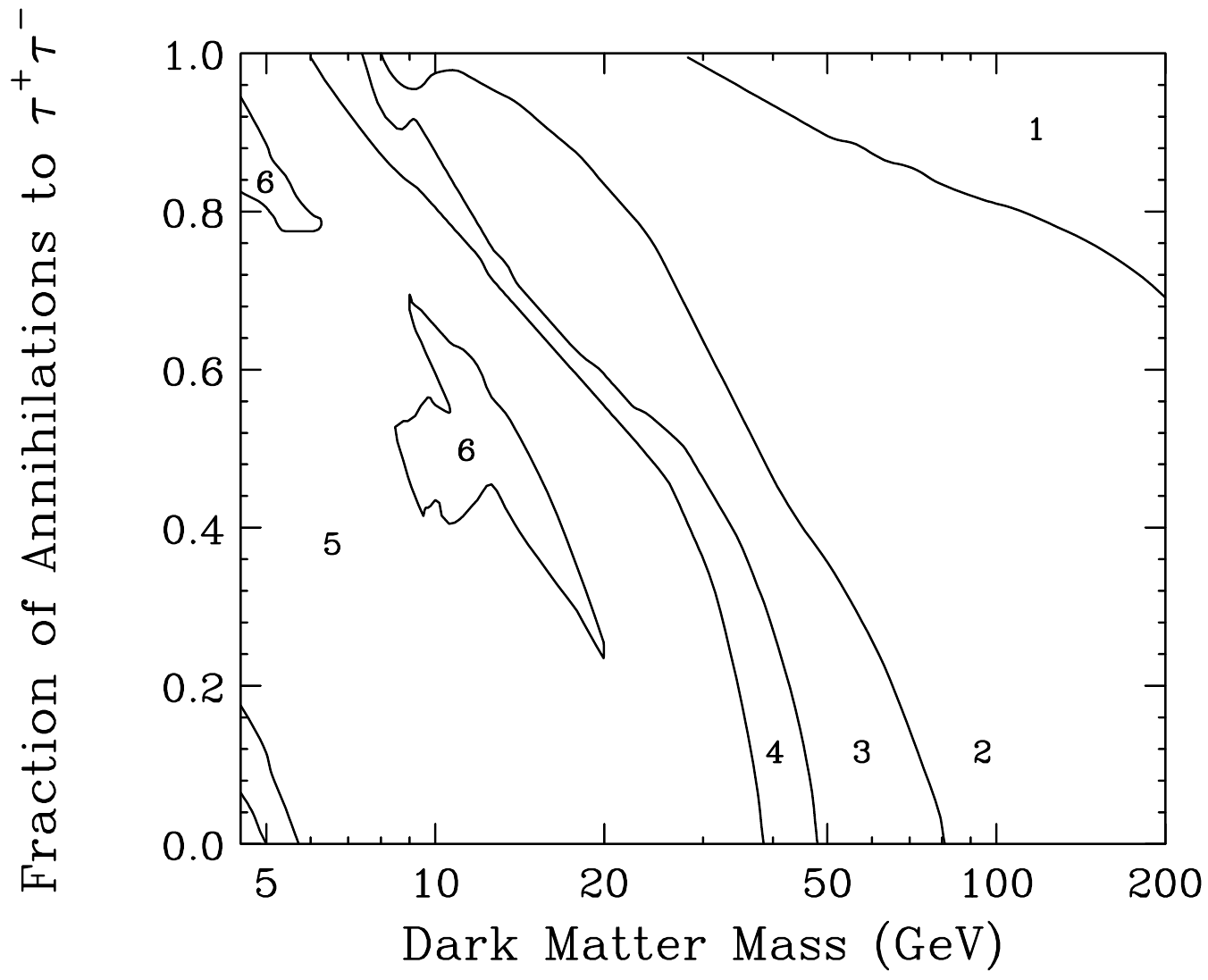}}
\caption{The number (out of a total of nine) of high-latitude ($|b|>60^{\circ}$), high-significance ($>5\sigma$), non-variable, unidentified sources contained in the Fermi Second Source Catalog (2FGL) with a spectrum consistent with originating from dark matter annihilations, for a range of dark matter masses and annihilation channels (the fraction of annihilations to taus is shown, with the remaining fraction proceeding to $b\bar{b}$). If the dark matter is fairly light ($m_{\rm DM} \lsim 40$ GeV), its annihilations in nearby subhalos could account for the majority of the observed high-latitude unidentified gamma-ray sources.}
\label{contours}
\end{center}
\end{figure*}

The Fermi-LAT collaboration has recently released its second source catalog (2FGL), which describes the observed characteristics of 1873 gamma ray sources. The 2FGL is based on 24 months of data and improves upon the previous Fermi source catalog by making use of the improved Pass 7 event selection, as well as a new high-resolution model of the diffuse Galactic and isotropic emissions~\cite{2catalog}. Unlike the first Fermi source catalog, the 2FGL reports flux measurements in 5 energy bands (100-300 MeV, 300-1000 MeV, 1-3 GeV, 3-10 GeV, and 10-100 GeV)~\cite{2catalog}, rather than as a single power-law fit~\cite{catalog}.

Of the 1873 sources contained in the 2FGL, 576 have not been identified or associated with counterparts at other wavelengths. Of these 576 unidentified sources, 397 have been detected at greater than 5$\sigma$ significance, and do not exhibit variability.

In Fig.~\ref{catalog} we show the distribution of the high-significance ($>$$5\sigma$), non-variable sources with Galactic latitude. In the upper frames, we show distribution of identified/associated sources. As expected, the Galactic sources (supernova remnants, pulsars, globular clusters, pulsar wind nebulae) are highly concentrated around the Galactic Plane. The extragalactic sources (active galactic nuclei, BL Lac objects, flat spectrum radio quasars, normal galaxies, radio galaxies, Seyfert galaxies, starburst galaxies) are more flatly distributed with $b$. As shown in the lower frames, the unidentified sources are even more heavily concentrated around the Galactic Plane. However, this effect is very likely attributable, at least in part, to the relative incompleteness of source catalogs at other wavelengths at low latitudes. In particular, optical identifications of radio sources are impeded by interstellar obscuration, leading many radio sources to remain unidentified. This bias leads to there being a large fraction of unidentified gamma ray sources at low latitude~\cite{2catalog}.

As observable dark matter subhalos are expected to be nearly isotropically distributed, we focus our search on high latitude regions of the sky where relatively few Galactic sources will be present. In particular, we begin by focusing on the nine high-significance, non-variable, unidentified sources with $|b|>60^{\circ}$. The spectra of these sources, as reported in the 2FGL, are shown in Fig.~\ref{nine}.  The highest significance and best measured of these sources (2FGL J2339.6-0532) exhibits a spectrum which peaks sharply at 1-3 GeV. In contrast, the source 2FGL J1129.5+3758 exhibits a very power-law like behavior. Most of the other nine sources are only measured in one or two energy bins, and thus provide us with limited spectral information.

To assess how many (if any) of these unidentified sources could potentially be dark matter subhalos, we compare their measured spectra to that predicted from dark matter with a range of masses and annihilation channels. In Fig.~\ref{contours} we show the number of high-latitude ($|b|>60^{\circ}$), high-significance ($>$$5\sigma$), non-variable, unidentified sources contained within the 2FGL with a spectrum that is consistent with originating from dark matter annihilations, for a range of dark matter masses and annihilation channels (the fraction of annihilations to taus is shown, with the remaining fraction proceeding to $b\bar{b}$).\footnote{The criteria we apply to consider a given source to be ``well-fit'' by the spectrum of a given dark matter model is $\chi^2 < 7.77$ (over 5-1 degrees of freedom), which was chosen to retain 90\% of sources with that spectrum.} 

If the dark matter particle is relatively heavy ($m_{\rm DM} \gsim 50$--$100$ GeV), only a small fraction of the nine candidate sources could potentially be dark matter subhalos. This is in large part due to the fact that none of these sources appear to be very luminous at energies above 10 GeV, making it difficult to accommodate the spectrum predicted for a heavy dark matter particle (see also, Ref.~\cite{Zechlin:2011wa}). In contrast, if the dark matter is somewhat light ($m_{\rm DM} \lsim 40$ GeV), its annihilations in nearby subhalos could account for the majority of the observed high-latitude unidentified gamma-ray sources.

Comparing these results to the estimates of the number of observable subhalos made in Sec.~\ref{theory}, we should not be surprised that we have not not identified many high-mass dark matter subhalo candidates in the 2FGL. A 200 GeV dark matter particle annihilating to $b\bar{b}$, for example, is expected to lead to no more than about a dozen observable halos, even for our optimistic assumptions and for a cross section that saturates the dwarf limit (see Fig.~\ref{number}). This corresponds to no more than about 1.6 observable subhalos at high latitudes ($|b|>60^{\circ}$). On the other hand, a 10-20 GeV dark matter particle could provide us with up to $\sim$80 observable subhalos, corresponding to about 11 at high-latitudes.

We also note that a similar fraction of the unidentified sources at mid-latitudes ($60^{\circ}>|b|>30^{\circ}$) are compatible with the spectrum from a 10-20 GeV dark matter particle (between 20 and 32 out of 58 sources provide a good fit for $m_{\rm DM}$ in the range of 10 to 20 GeV and for any combination of annihilations to $b\bar{b}$ and $\tau^+ \tau^-$). A significantly smaller fraction of low latitude sources provide a good fit. We caution, however, that this distinction could plausibly result from biases associated with the varying diffuse background.

In the following section, we consider relatively light dark matter particles further within this context, focusing on dark matter particles with a mass and dominant annihilation channels that can accommodate the gamma ray signal observed from the Galactic Center, as well as the observed spectra of the Milky Way's radio filaments and the microwave haze observed by WMAP.



\section{The Dark Matter Model Motivated by the Galactic Center}
\label{gc}

The high density of dark matter predicted to be present in the region surrounding the Galactic Center make it a very promising target for indirect searches for dark matter. The spectrum and morphology of the gamma rays observed from this region by the FGST have been studied and found to be consistent with that predicted from an approximately 7-12 GeV dark matter particle which annihilates mostly to leptons (in particular, to $\tau^+ \tau^-$) possibly along with a smaller fraction of annihilations proceeding to $b\bar{b}$ or other hadronic final states~\cite{Hooper:2011ti,Hooper:2010mq}. This scenario is further supported by the peculiar spectral radio features observed from many of the Milky Way's non-thermal radio filaments~\cite{Linden:2011au} and by the Inner Galaxy's microwave haze~\cite{Hooper:2010im}, each of which can be explained by annihilating dark matter with the same mass, annihilations channels, cross section, and halo profile as are required to produce the gamma ray signal from the Galactic Center. Furthermore, a number of direct detection experiments (DAMA/LIBRA~\cite{dama}, CoGeNT~\cite{cogent} and CRESST-II~\cite{cresst}) have reported signals consistent with $\sim$10 GeV dark matter particles~\cite{Kelso:2011gd}.

To consider a concrete example, we consider a 10 GeV dark matter particle with a total annihilation cross section of $\sigma v\ = 3 \times 10^{-26}$ cm$^3$/s, and which annihilates to $\tau^+ \tau^-$, $\mu^+ \mu^-$, and $e^+ e^-$ 30\% of the time each, and 10\% of the time to $b\bar{b}$ (corresponding to a fraction to $\tau^+ \tau^-$ of 0.75 in Fig.~\ref{contours}, as annihilations to muons and electrons contribute insignificantly to the total gamma ray spectrum). Such a dark matter candidate would be capable of producing the observed Galactic Center gamma rays, as well as the signals from radio filaments and the microwave haze. For our central parameter estimates, this model leads to a prediction of 3.1 subhalos across the sky that are detectable by the FGST, or 43 if our more optimistic parameters are used. This corresponds to approximately 0.4 (5.8) observable subhalos at $|b|>60^{\circ}$ for our central (optimistic) parameter choices. Comparing this to the results shown in Fig.~\ref{contours}, we see that 5 out of 9 of the unidentified, high-latitude, high-significance, non-variable sources in the 2FGL can be fit by this dark matter model, which is within the predicted range predicted in this scenario. While it is not possible at this time to conclude that these sources are in fact dark matter subhalos, rather than AGN or other astrophysical objects, the similarity between the Galactic Center gamma ray spectrum and the spectra of these unidentified sources provides a strong motivation to study these objects further, at both gamma ray and other other wavelengths. If no X-ray, radio, optical or other counterparts to any of these sources can be found, it would strengthen the case that they may be dark matter subhalos.

\section{Summary and Conclusions}
\label{conclusions}

Dark matter halos are predicted to contain large numbers of smaller and dense subhalos. In many typical dark matter models, annihilations take place within these subhalos at a rate sufficient for them to appear as bright and approximately point-like gamma ray sources. A 10 GeV (100 GeV) dark matter particle with a roughly thermal annihilation cross section ($\sigma v\sim 3\times 10^{-26}$ cm$^3$/s) is expected to provide approximately 1 to 200 (up to $\sim$10) subhalos that are observable to the Fermi Gamma Ray Space Telescope. 

The Fermi-LAT Second Source Catalog contains 576 sources which have not been identified or associated with counterparts at other wavelengths. Of these sources, 397 have been detected at high-significance ($>$$5\sigma$) and do not exhibit variability. In studying the spectrum of these sources, we find that a significant fraction of those found at high and mid-Galactic latitudes possess spectra that are consistent with that predicted from the annihilations of a relatively light dark matter particle ($m_{\rm DM}\sim 5-40$ GeV). If dark matter annihilations are responsbility for these gamma ray sources, it suggests that the Second Fermi Catalog contains a population of several tens of dark matter subhalos, in good agreement with the number expected for a thermal relic ($\sigma v\sim 3\times 10^{-26}$ cm$^3$/s).

To further test the hypothesis that a sizable population of dark matter subhalos exists within the Fermi-LAT Second Source Catalog, it is essential to search for counterparts of these sources in optical, X-ray, or other wavelengths.  If no such emission can be found, it would strengthen the case that these sources may be dark matter subhalos. Additionally, as the Fermi Telescope collects more data, it will increasingly improve their measurements the spectra of these sources, and become more sensitive to any angular extention they may exhibit.

\bigskip

{\it Acknowledgements:} MB and DH are supported by the US Department of Energy, including grant DE-FG02-95ER40896.

\end{document}